\documentclass{appolb}
\usepackage{graphicx}
\usepackage{amsmath,amsfonts,mathtools}
\usepackage{amssymb}

\begin{document}
\title{Thermal leptogenesis in minimal unified models%
\thanks{Presented at the ``V4-HEP 3 - Theory and Experiment in High Energy Physics" workshop in Prague, October 3 2024}%
}
\author{Michal Malinsk\'{y}
\address{Institute of Particle and Nuclear Physics,
  Faculty of Mathematics and Physics,
  Charles University in Prague, V Hole\v{s}ovi\v{c}k\'ach 2,
  180 00 Praha 8, Czech Republic}
}
\maketitle
\begin{abstract}
We review the status of thermal leptogenesis in the minimal $SU(5)\times U(1)$ and $SO(10)$ unified models under the assumption that the leptonic asymmetry generated in the out-of-equilibrium decays of heavy Majorana neutrinos (and its subsequent conversion into baryons via sphalerons) constitutes the primary source of baryon asymmetry of the Universe. In both cases, leptogenesis is shown to provide a strong extra constraint on the flavour structure of the model under consideration, leading to interesting and potentially testable phenomenological effects.     
\end{abstract}
%%%%%%%%%%%%%%%%%%%%%%%%%%%%%%%%%%%%%%%%%%%%%%%%%%%%%%%%%%%%%%%%%%%%%%%%%%%%%%  
\section{Introduction}
With the advent of the next generation of ultra-large-scale detectors such as Hyper-K~\cite{Hyper-Kamiokande:2018ofw} and DUNE~\cite{DUNE:2016hlj} a new round of proton (and bound neutron) decay searches with unprecedented sensitivity is being prepared. Such a campaign would naturally benefit from a good and consensual theoretical picture providing, at least, hints on the relevant lifetime(s) and, optimally, even branching ratios.  

This, however, is not what is readily available even within the most conservative approach to this kind of physics, namely, the Grand unified theories (or their descendants). Most of the models on the market, indeed, suffer from at least one (if not all) of the following drawbacks, crippling their capacity to keep at least the main theoretical uncertainties of their predictions under reasonable control:
\begin{enumerate}

\item Bad grip on the location of the unification scale which defines the masses of the heavy mediators behind the $d=6$ effective baryon and lepton number ($B$ and $L$) violating Standard Model (SM) operators. This is typically due to i) insufficient information on the high-energy spectrum of the model, owing to frequent technical difficulties with its detailed calculation, especially in case of ``baroque''constructions, ii) insufficient accuracy of the calculation of  relevant $\beta$-functions, often related to point i), iii) the relatively small hierarchy between the unification scale $M_U$ (typically in the $10^{16}$~GeV ballpark) and the Planck scale $M_{\rm Pl}$ (at around $10^{18}$ GeV) which tends to break down the traditionally adopted renormalizable approximation, especially when it comes to precision calculations. 

\item Limited information on the structure of the $B$ and $L$ violating vector and scalar currents governing the proton decay amplitudes. This is namely due to the fact that the available low-energy flavour data (i.e. quark and lepton mass spectra and the mixing angles and CP phases parametrising the CKM and PMNS matrices) are typically not enough to fully reconstruct all the relevant charged-current flavour structures thereof. Hence, besides the most minimal models with simplest Yukawa sectors, there is practically no chance to go much beyond qualitative predictions concerning, at best, very specific $p$-decay channels (exploiting, for instance, their partial blindness to such details in specific channels like e.g. $p\to \pi^+\overline{\nu}$, where the summation over the final state neutrino flavours and the unitarity of the mixing matrices in the vector currents often simplifies the matters significantly).    

\item Bad or very limited grip on low-scale thresholds which, in certain types of models, tend to plague the desired gauge and/or Yukawa unification analysis of points 1) and 2). This is the typical case of supersymmetric GUTs where even the most conservative SUSY breaking schemes (in absence of any direct experimental constraints) do not provide sufficient information to pinpoint the details of the low-scale spectrum of the theory.
 
\end{enumerate}

Interestingly, several notable exceptions challenge this rather grim outlook. Among these, two minimal unified scenarios stand out, namely, the minimal flipped $SU(5)$~\cite{ArbelaezRodriguez:2013kxw} (sometimes even referred to as the minimal renormalizable model of baryon and lepton number violation) and the minimal potentially realistic $SO(10)$ unification~\cite{Chang:1984qr,Deshpande:1992au,Bertolini:2009es} which is arguably the best chance to overcome the hurdles above in the genuine GUT context. In both these cases, one entertains a relatively good grip on the high-scale spectrum, the most prominent uncertainties related to the proximity of $M_U$ and $M_{\rm Pl}$ are under reasonable control, and their Yukawa sector is simple enough to admit rather detailed numerical simulations.

In the current study, we aim namely on these two settings, focusing on their capacity to account for a sufficiently large baryon-antibaryon asymmetry generated in the early Universe by the thermal leptogenesis mechanism~\cite{Fukugita:1986hr}, corresponding to a large-enough baryon-to-photon number density ratio $\eta_B$ (in the $6\times 10^{-10}$ ballpark). In both cases we shall for simplicity assume that leptogenesis is the primary source of the asymmetry, i.e. we shall neglect the direct net $B$ production at the unification scale (usually attributed to $B$-violating decays of heavy scalar triplets). 
This extra piece of information typically translates into additional constraints on their Yukawa structure, which, in the minimal models, tends to be already rather overloaded by the basic need to accommodate the low-energy quark and lepton mass spectra and their mixings. As we shall see, even with a single such extra constraint added to the play, interesting insights with potentially far reaching consequences can be obtained.                      
%%%%%%%%%%%%%%%%%%%%%%%%%%%%%%%%%%%%%%%%%%%%%%%%%%%%%%%%%%%%%%%%%%%%%%%%%%%%%%   
\section{Thermal leptogenesis in the minimal $SU(5)\times U(1)$ unification}
\begin{figure}[htb]
\centerline{%
\includegraphics[width=4cm]{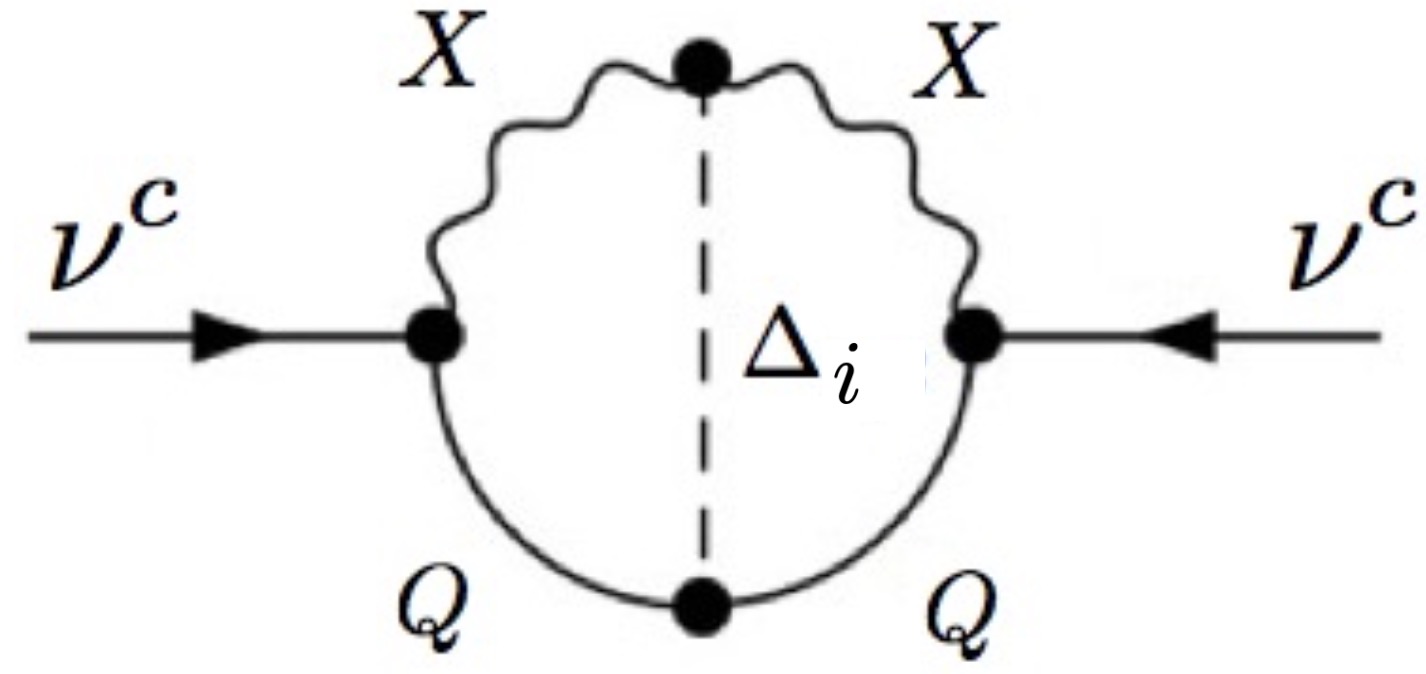}\hspace{1cm}
\includegraphics[width=4cm]{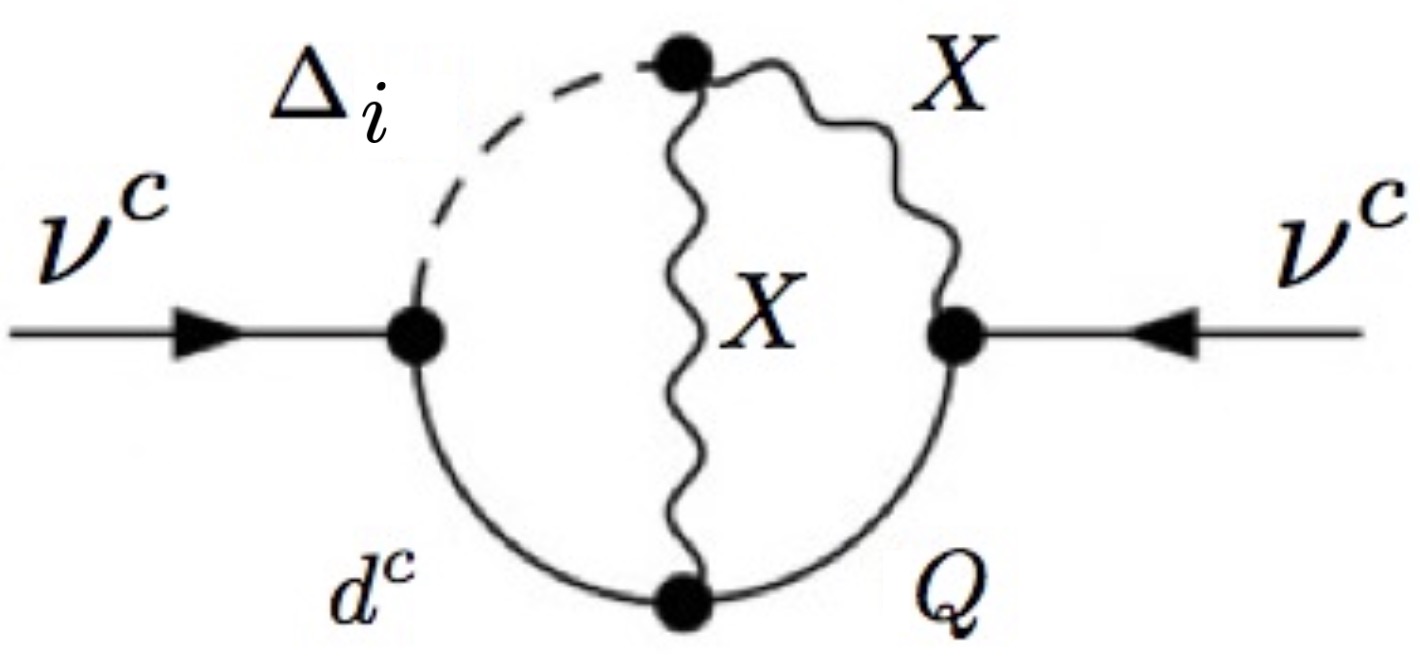}}
\caption{Two-loop Feynman diagrams giving rise to a radiatively generated Majorana mass for RH neutrinos in the minimal flipped $SU(5)$ model of Ref.~\cite{ArbelaezRodriguez:2013kxw}. $Q$, $d^c$ and $\nu^c$ denote the SM quark doublets, down-type quark singlets and the RH neutrinos, while $X$ and $\Delta_i$ are the unification-scale vector bosons and coloured scalar triplets, respectively.}
\label{Fig:graphs}
\end{figure}

The minimal flipped $SU(5)$ unified theory~\cite{ArbelaezRodriguez:2013kxw,Harries:2018tld,Fonseca:2023per} is a variant of the original scheme~\cite{Barr:1981qv} in which one does not need to resort to a large (i.e. dimension 50) scalar representation to devise the Majorana mass term for RH neutrinos; rather than that, the seesaw scale in this setting is generated radiatively at two loops; see Fig~\ref{Fig:graphs} (and also Ref.~\cite{Leontaris:1991mq}). 
Due to this effect, the model features a very constrained Yukawa sector in which the Dirac neutrino mass matrix is strongly correlated with that of the up-type quarks $M_\nu^D=M_u^T$, the structure of the RH neutrino Majorana mass matrix $M_\nu^M$ corresponds to a simple combination of Yukawa couplings governing the masses of down-type quarks etc. As a consequence, the ratios of all 2-body proton decay partial widths (and, hence, their branching ratios) are in principle calculable in terms of a single a-priori unknown unitary matrix $U_\nu$ (or its compound $U_\ell^L\equiv V_{\rm PMNS}U_\nu$) which diagonalizes the light Majorana neutrino mass matrix $m_{\nu}=U_\nu^\dagger D_\nu U_\nu^*$ in the basis where up-quark masses are diagonal; for instance (see Ref.~\cite{ArbelaezRodriguez:2013kxw}): 
\begin{align}
\label{form1}
\Gamma(p\to \pi^0 \mu^+)& =\frac{1}{2}|(V_{CKM})_{11}|^2|(U^L_\ell)_{21}|^2 \Gamma(p\to \pi^+\overline{\nu})\,,\\
\label{form2}
\Gamma(p\to K^0 e^+)& =\frac{C_3}{C_1}|(V_{CKM})_{12}|^2|(U^L_\ell)_{11}|^2 \Gamma(p\to \pi^+\overline{\nu})\,,
\end{align}
etc.; here $C_{1,3}$ are calculable chiral factors and $V_{CKM}$ is the Cabibbo-Kobayashi-Maskawa matrix.

Remarkably, perturbativity and unification constraints, together with the need to accommodate the neutrino oscillation data, strongly constrain the consistent shapes of $U_\nu$ on their own, cf.~\cite{ArbelaezRodriguez:2013kxw}. Hence, the additional requirement to reproduce the correct $\eta_B$ turns out to be especially powerful in  this scenario. 

\begin{figure}[htb]
\centerline{%
\includegraphics[width=7.3cm]{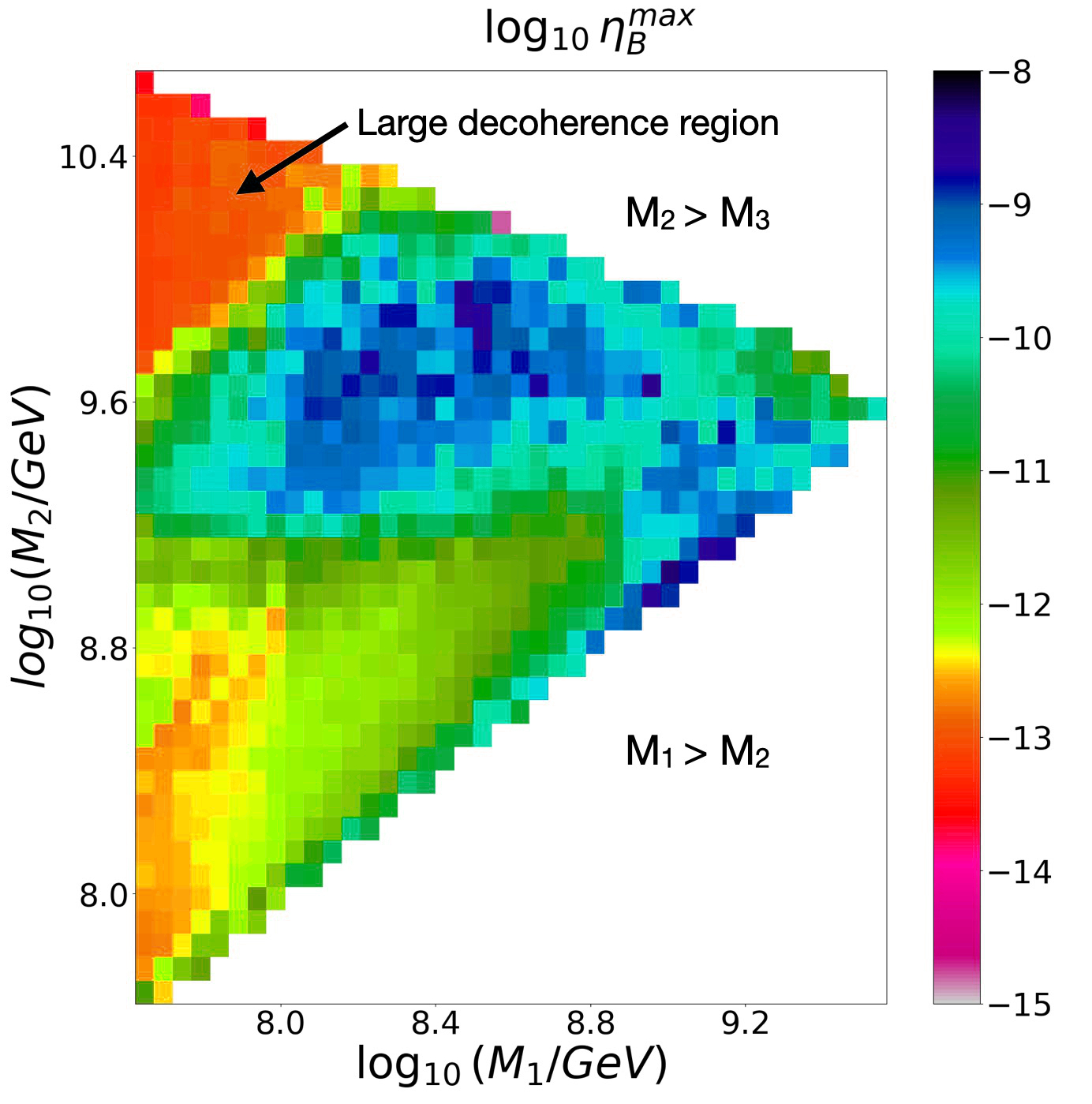}
\includegraphics[width=7cm]{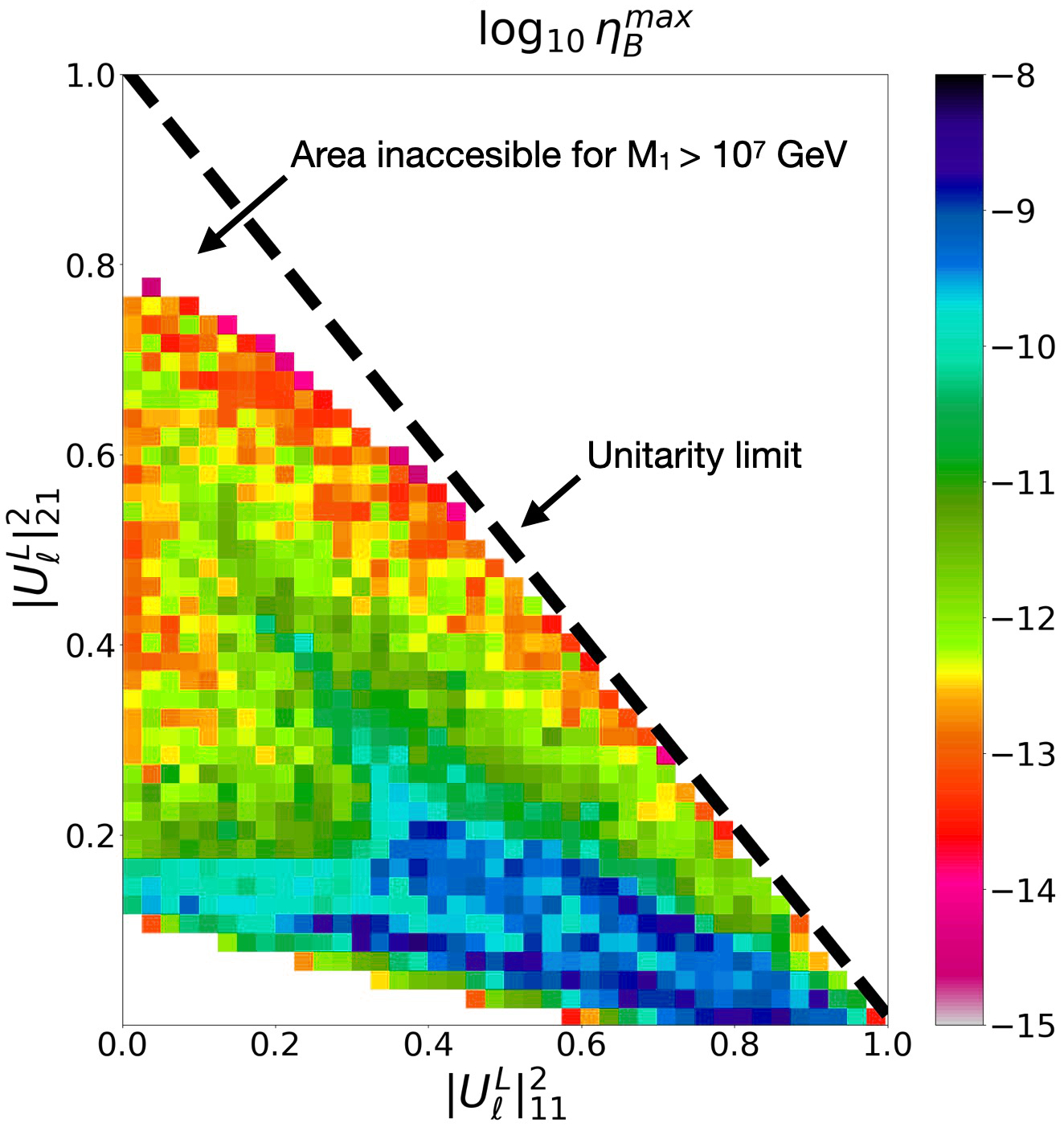}}
\caption{Maximum levels of attainable baryon asymmetry displayed along two different cuts of the parameter space of the minimal $SU(5)\times U(1)$ model for the normal hierarchy case and the mass of the lightest neutrino at the level of $m_0\sim 1$~meV. Acceptable areas (yielding $\eta_B\gtrsim 6\times 10^{-10}$) correspond to dark blue regions. From the left panel, one can determine the  limits on the masses of the lighter two heavy neutrinos in whose decays the leptonic CP asymmetry is typically generated; in the right panel, bounds on elements of the unitary matrix $U_\ell^L$ steering proton decay in different  flavour channels are shown, cf. formulae~(\ref{form1}) and~(\ref{form2}).}
\label{Fig:plots}
\end{figure}

The full numerical analysis of the impact of this extra piece of information on the shape of the parameter space of the model has been performed in Ref.~\cite{Fonseca:2023per}. For that sake, the ULYSSES package~\cite{Granelli:2020pim} has been employed together with a dedicated scanning algorithm which was focusing on the most promising patches thereof. Three different regimes in which large-enough asymmetry was attainable have been identified, namely\footnote{Note that the three different regimes can be to some extent identified with the three dark domains visible on the left panel of Fig.~\ref{Fig:plots}.}:  
\begin{itemize}
\item A regime in which the $L$-asymmetry production is dominated by decays of $N_1$ (the lightest heavy neutrino mass eigenstates) due to an enhancement in the corresponding CP asymmetry competing with a significant washout.     
\item A regime in which the final asymmetry created mainly in decays of $N_2$ is not entirely washed out by $N_1$ driven dynamics due to a suppression of the relevant decoherence effects.     
\item A regime in which the final asymmetry is again dominated by $N_2$ decays, but this time in a mode in which the $N_2$-associated washout is strongly suppressed (and the decoherence effects are again kept under control).     
\end{itemize}
Remarkably, in none of these cases a large-enough $\eta_B$ was obtained for the mass of the lightest active neutrino above about 0.03 eV! This, in turn, provides an upper limit on the absolute neutrino mass scale and makes the scheme potentially testable in the current and near future beta-decay facilities like KATRIN~\cite{KATRIN:2022ayy}.        

%%%%%%%%%%%%%%%%%%%%%%%%%%%%%%%%%%%%%%%%%%%%%%%%%%%%%%%%%%%%%%%%%%%%%%%%%%%%%%   
\section{Thermal leptogenesis in the minimal $SO(10)$ unification}
The minimal $SO(10)$ GUT we shall be concerned with in this section is an old framework based on Refs.~\cite{Chang:1984qr,Deshpande:1992au} which, however, had been abandoned as a candidate for a potentially realistic theory shortly after its conception, see~\cite{Buccella:1980qb,Yasue:1980fy}. This was mainly due to the absence of a potentially realistic symmetry breaking pattern in the patches of the parameter space where a non-tachyonic scalar spectrum is supported. However, as shown in~\cite{Bertolini:2009es}, this was a mere artefact of the original tree-level approach and the model is potentially consistent at one loop~\cite{Jarkovska:2021jvw}, albeit rather difficult to handle in necessary detail.

Again, the model in its most minimal version features a strongly constrained Yukawa sector which has been subject to many studies (see e.g. \cite{Joshipura:2011nn,Dueck:2013gca,Altarelli:2013aqa,Babu:2015bna,Babu:2016bmy,Ohlsson:2019sja}) concerning mainly its capacity to accommodate the low-energy quark and lepton masses and mixings. Recently, such attempts have been augmented with the extra information from $\eta_B$ (see e.g.~\cite{Babu:2024ahk}) and the results are encouraging - large-enough $\eta_B$ is typically compatible with the general shape of the Yukawa sector required by the need to accommodate the SM fermionic data.   

In this paper, we present a small subset of preliminary results of an independent study of this kind~\cite{inpreparation} in which the stochastic version of the differential evolution algorithm (see for instance~\cite{diffevolalg}) optimized for a fast exploration of the potentially realistic Yukawa sector patterns has been, like in the flipped $SU(5)$ case above, combined with the powers of the ULYSSES package~\cite{Granelli:2020pim}. 

The most interesting observation here is perhaps the capacity of the fitter to assume global $\chi^2$ values well below 10; with almost 20 fitted parameters this indeed indicates a good compatibility of the minimal renormalizable $SO(10)$ Yukawa structure with the low-energy data. Remarkably, the  values of the lepton-sector Dirac CP phase in the best fit points fall rather consistently into the 3rd and 4th quadrants, aligning well with the range indicated by e.g. the latest T2K fits~\cite{T2K:2023smv}. At the same time, $\eta_B$ close to the experimental value can also be attained. 

\begin{figure}[htb]
\centerline{%
\includegraphics[width=6.5cm]{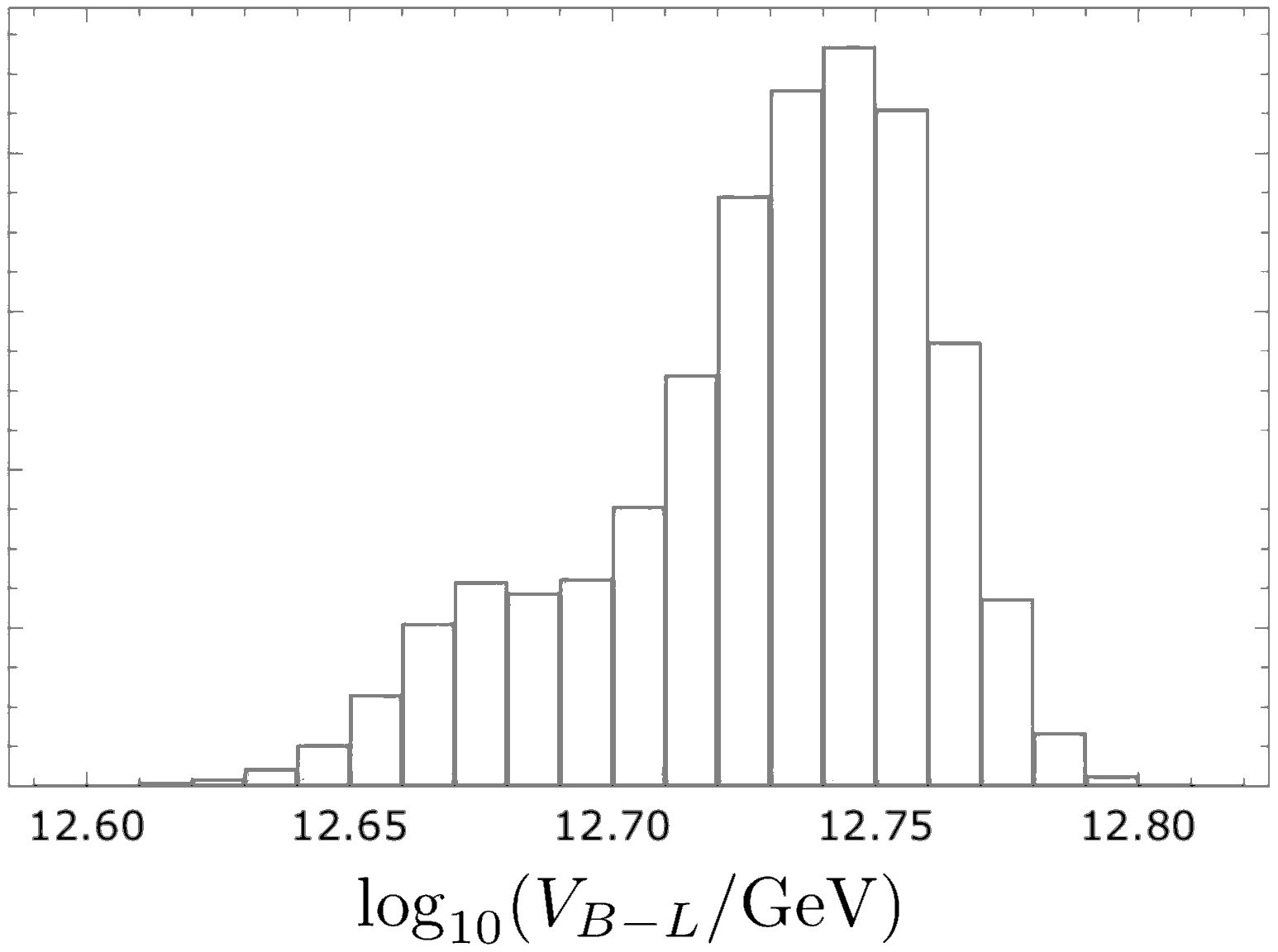}}
\caption{The $B-L$ breaking scale in the minimal $SO(10)$ GUT determined solely from the requirement of the compatibility of the flavor structure of the model with the measured value of the baryon asymmetry (assumed to be dominated by thermal  leptogenesis); no other constraints have been used in producing the plot. Remarkably enough, the scale of the $U(1)_{B-L}$  breaking VEV automatically falls into the vicinity of the region favoured by the detailed gauge coupling unification analysis~\cite{Jarkovska:2021jvw}.}
\label{Fig:F2}
\end{figure}

Another interesting feature of the presented sample of the numerical fits of~\cite{inpreparation} is the minimal $SO(10)$ model's capacity to localize the position of the $B-L$ breaking scale solely from the low-energy flavour data augmented with the extra constraint from $\eta_B$, without ever referring to the traditional analysis of the running gauge couplings. While the uncertainties due to this negligence (for instance, the use of a simplified threshold-free estimate of the value of the unified gauge coupling etc.) are under reasonable control, the $B-L$ breaking VEV of the relevant $SU(2)_R$ scalar triplet turns out to be in the $10^{12.5}$~GeV ballpark (see Fig.~\ref{Fig:F2}), not far from the range favoured by the $\omega_{\rm BL}\to 0$ scenario studied in detail in~\cite{Jarkovska:2021jvw}.

\section{Conclusions}
In this proceedings contribution we have briefly presented the (partly preliminary) results of two studies (one published, the other one still in preparation phase) attempting to exploit the observed baryon asymmetry of the Universe as an additional constraint imposed on the flavour structure of two specific minimal unified settings. In case of the minimal $SU(5)\times U(1)$ model, an interesting upper limit on the mass of the lightest active neutrino has been obtained. The preference of the negative value (i.e. the 3rd or 4th quadrant) for the leptonic Dirac CP phase in the low-$\chi^2$ fits of the Yukawa structure of the minimal potentially realistic $SO(10)$ GUT is also quite intriguing. Hence, the extra information from cosmology tends to enhance the predictive power of these models, rendering both of them potentially testable at the existing or near-future facilities.

%\bibliographystyle{h-physrev5}
%\bibliography{bibliography}

\begin{thebibliography}{10}

\bibitem{Hyper-Kamiokande:2018ofw}
Hyper-Kamiokande, K.~Abe {\em et~al.},
\newblock (2018), arXiv:1805.04163.

\bibitem{DUNE:2016hlj}
DUNE, R.~Acciarri {\em et~al.},
\newblock (2016), arXiv:1601.05471.

\bibitem{ArbelaezRodriguez:2013kxw}
C.~Arbel\'aez~Rodr\'\i{}guez, H.~Kole\v{s}ov\'a, and M.~Malinsk\'y,
\newblock Phys. Rev. D {\bf 89}, 055003 (2014), arXiv:1309.6743.

\bibitem{Chang:1984qr}
D.~Chang, R.~N. Mohapatra, J.~Gipson, R.~E. Marshak, and M.~K. Parida,
\newblock Phys. Rev. D {\bf 31}, 1718 (1985).

\bibitem{Deshpande:1992au}
N.~G. Deshpande, E.~Keith, and P.~B. Pal,
\newblock Phys. Rev. D {\bf 46}, 2261 (1993).

\bibitem{Bertolini:2009es}
S.~Bertolini, L.~Di~Luzio, and M.~Malinsky,
\newblock Phys. Rev. D {\bf 81}, 035015 (2010), arXiv:0912.1796.

\bibitem{Fukugita:1986hr}
M.~Fukugita and T.~Yanagida,
\newblock Phys. Lett. B {\bf 174}, 45 (1986).

\bibitem{Harries:2018tld}
D.~Harries, M.~Malinsk\'y, and M.~Zdr\'ahal,
\newblock Phys. Rev. D {\bf 98}, 095015 (2018), arXiv:1808.02339.

\bibitem{Fonseca:2023per}
R.~Fonseca, M.~Malinsk\'y, V.~Mi\v{r}\'atsk\'y, and M.~Zdr\'ahal,
\newblock Phys. Rev. D {\bf 110}, 015030 (2024), arXiv:2312.08357.

\bibitem{Barr:1981qv}
S.~M. Barr,
\newblock Phys. Lett. B {\bf 112}, 219 (1982).

\bibitem{Leontaris:1991mq}
G.~K. Leontaris and J.~D. Vergados,
\newblock Phys. Lett. B {\bf 258}, 111 (1991).

\bibitem{Granelli:2020pim}
A.~Granelli, K.~Moffat, Y.~F. Perez-Gonzalez, H.~Schulz, and J.~Turner,
\newblock Comput. Phys. Commun. {\bf 262}, 107813 (2021), arXiv:2007.09150.

\bibitem{KATRIN:2022ayy}
KATRIN, M.~Aker {\em et~al.},
\newblock J. Phys. G {\bf 49}, 100501 (2022), arXiv:2203.08059.

\bibitem{Buccella:1980qb}
F.~Buccella, H.~Ruegg, and C.~A. Savoy,
\newblock Phys. Lett. B {\bf 94}, 491 (1980).

\bibitem{Yasue:1980fy}
M.~Yasue,
\newblock Phys. Rev. D {\bf 24}, 1005 (1981).

\bibitem{Jarkovska:2021jvw}
K.~Jarkovsk\'a, M.~Malinsk\'y, T.~Mede, and V.~Susi\v{c},
\newblock Phys. Rev. D {\bf 105}, 095003 (2022), arXiv:2109.06784.

\bibitem{Joshipura:2011nn}
A.~S. Joshipura and K.~M. Patel,
\newblock Phys. Rev. D {\bf 83}, 095002 (2011), arXiv:1102.5148.

\bibitem{Dueck:2013gca}
A.~Dueck and W.~Rodejohann,
\newblock JHEP {\bf 09}, 024 (2013), arXiv:1306.4468.

\bibitem{Altarelli:2013aqa}
G.~Altarelli and D.~Meloni,
\newblock JHEP {\bf 08}, 021 (2013), arXiv:1305.1001.

\bibitem{Babu:2015bna}
K.~S. Babu and S.~Khan,
\newblock (2015), arXiv:1507.06712.
%%CITATION = ARXIV:1507.06712;%%

\bibitem{Babu:2016bmy}
K.~S. Babu, B.~Bajc, and S.~Saad,
\newblock JHEP {\bf 02}, 136 (2017), arXiv:1612.04329.

\bibitem{Ohlsson:2019sja}
T.~Ohlsson and M.~Pernow,
\newblock JHEP {\bf 06}, 085 (2019), arXiv:1903.08241.

\bibitem{Babu:2024ahk}
K.~S. Babu, P.~Di~Bari, C.~S. Fong, and S.~Saad,
\newblock JHEP {\bf 10}, 190 (2024), arXiv:2409.03840.

\bibitem{inpreparation}
M.~Malinsk\'y and D.~Star\'{y},
\newblock in preparation.

\bibitem{diffevolalg}
M.~Georgioudakis and V.~Plevris,
\newblock Frontiers in Built Environment {\bf 6}, 102 (2020).

\bibitem{T2K:2023smv}
T2K, K.~Abe {\em et~al.},
\newblock Eur. Phys. J. C {\bf 83}, 782 (2023), arXiv:2303.03222.

\end{thebibliography}

\end{document}